\documentclass[aps,prb,twocolumn,showpacs,amsmath,amssymb]{revtex4-1}
\usepackage{graphicx}
\usepackage{bm}
\usepackage{amsmath}

\newcommand{\be}{\begin{equation}}
\newcommand{\ee}{\end{equation}}
\newcommand{\bea}{\begin{eqnarray}}
\newcommand{\eea}{\end{eqnarray}}
\newcommand{\up}{\uparrow}
\newcommand{\down}{\downarrow}
\newcommand{\bwt}{\begin{widetext}}
\newcommand{\ewt}{\end{widetext}}
\newcommand{\ham}{\mathcal{H}}

\newcommand{\ra}{\rangle}
\newcommand{\la}{\langle}
\newcommand{\bsb}{\begin{subarray}}
\newcommand{\esb}{\end{subarray}}
\newcommand{\largem}{\!\!}

\newcommand{\eins}{\mbox{$1 \hspace{-1.0mm} {\bf l}$}}

\newcommand{\vecv}[2]{
\left(\largem
 \begin{tabular}{c}
  $#1$ \\
  $#2$
  \end{tabular}
  \largem
\right)
}

\newcommand{\vech}[2]{
\left(\largem
 \begin{tabular}{c}
  $#1$ \! $#2$
  \end{tabular}
  \largem
\right)
}

\newcommand{\mat}[4]{
\left(
\largem
 \begin{tabular}{cc}
  $#1$ & $#2$ \\
  $#3$ & $#4$
  \end{tabular}
  \largem
\right)
}

\begin{document}

\title{Superconducting  current and proximity effect in ABA and ABC multilayer graphene Josephson junctions}

\author{W.A. Mu\~noz}
\email{WilliamArmando.Munoz@uantwerp.be}
\author{L. Covaci}
\email{lucian@covaci.org}
\author{F.M. Peeters}
\email{Francois.Peeters@uantwerp.be}
\affiliation{Departement Fysica, Universiteit Antwerpen, Groenenborgerlaan 171, B-2020 Antwerpen, Belgium}

\date{\today}

\begin{abstract}
Using a numerical tight-binding approach based on the Chebyshev-Bogoliubov-de Gennes method we describe Josephson junctions made of multilayer graphene contacted by top superconducting gates. Both Bernal (ABA) and rhombohedral (ABC) stacking are considered and we find that the type of stacking has a strong effect on the proximity effect and the supercurrent flow. For both cases the pair amplitude shows a polarization between dimer and non-dimer atoms, being more pronounced for rhombohedral stacking. Even though the proximity effect in non-dimer sites is enhanced when compared to single layer graphene, we find that the supercurrent is  suppressed. The spatial distribution of the supercurrent shows that for Bernal stacking the current flows only in the top-most layers while for rhombohedral stacking the current flows throughout the whole structure.
\end{abstract}



\maketitle

%
The exceptional characteristics of carriers in a single graphene layer gives rise to unusual properties of superconductor-graphene junctions such as specular Andreev reflection \cite{beenakker_specular_2006} and finite superconducting current at the neutrality point \cite{titov_josephson_2006}. Although there is still no clear evidence of  the novel electron-hole conversion, a bipolar proximity-induced supercurrent has been detected in superconducting-graphene-superconducting Josephson junctions (JJ) \cite{heersche_bipolar_2007, du_josephson_2008,ojeda-aristizabal_tuning_2009,jeong_observation_2011} opening a new perspective for Josephson field transistors. 
In these devices, carrier density modulations by the gate voltage playes an important role in controlling the strength of the proximity effect and therefore the dissipationless current flowing through the junction.
It is also expected that other graphene structures show interesting properties when in contact with superconducting leads. In fact, Josephson junctions with non-superconducting few-layer graphite films have been the focus of experimental investigations \cite{shailos_proximity_2007,sato_different_2008, sato_gate-controlled_2008,kanda_dependence_2010}.
In most of the few preceding theoretical studies \cite{hayashi_superconducting_2010,takane_tunneling_2010,hayashi_theoretical_2010} the proximity-induced superconducting correlations in multilayer graphene were determined using analytical approximations where the electronic description was limited to parabolic energy bands near the Fermi energy. As a consequence the depth dependence of the order parameter was neglected and in some cases the superconducting pair diffusion was reduced to a 2-dimensional scenario, therefore ignoring any spread of the Cooper pairs among the different layers. Since all the experimental setups require top superconducting contacts, a calculation taking into account the depth dependence of the pair correlation is needed.

In this Rapid Communication we describe the 3-dimensional diffusion of Cooper pairs through a non-superconducting multilayer graphene junction connected to two top superconducting electrodes. The Josephson superconducting current is also studied by setting a phase gradient between the superconducting leads. We find significant differences between the two possible stacking orders, Bernal and rhombohedral. For junctions with Bernal stacking the supercurrent flows mostly through the two top-most layers while for junctions with rhombohedral stacking the current is weaker and spread throughout the whole multilayer.\\
\begin{figure}[ttt]
\includegraphics[width=0.95\columnwidth]{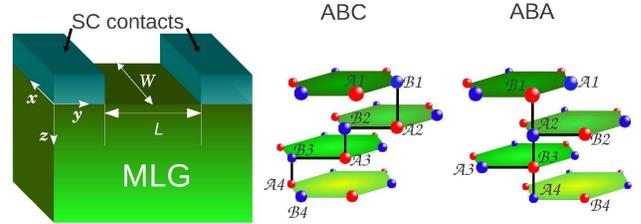}	
\caption{  \label{layout} (Color online) Schematic layout of a multilayer graphene JJ showing the superconducting (SC) leads which are separated by a distance $L$. $W$ corresponds to the width of the junction. On the right-hand side the layer arrangements in multilayer graphene shows the two different stacking configuration: ABC(rhombohedral) and ABA(Bernal).
}\label{layout}
\end{figure}

We consider a multilayer graphene junction, shown in Fig.~\ref{layout}, where the top layer is in contact with two superconducting leads separated by a distance $L$, which corresponds to the junction length. We adopt the non self-consistent  method employed in Ref.~[\onlinecite{covaci_superconducting_2011}] where three-dimensional superconducting leads were assumed to act as external reservoirs of Cooper pairs. In our case the superconducting contacts,  are single layer graphene with an intrinsic s-wave order parameter,  $\Delta_0=0.1t_0$, and a high doping level, $\mu_s$=0.6t, where $t_0$=2.8eV. The coupling between the graphene multilayer and the superconducting contacts is chosen such that there is a sizable proximity effect and the edge effects are minimal for the size of the junctions considered here. At the present stage, we assume that the inverse proximity effect on the superconducting contacts is negligible.

It is well known that the electronic properties of multilayer graphene depend strongly on the particular type of coupling between the graphene layers. Likewise, intrinsic electronic correlations have been shown to behave differently according to the stacking configuration \cite{munoz_tight-binding_2013}. A question that arises naturally is: how different is the superconducting pairing diffusion under the contacts and across the junction for different number of layers and stacking configurations?

Based on this motivation, we perform calculations considering the two most stable interlayer stackings found in multilayer graphene:  ABA or Bernal, and ABC or rhombohedral. Bernal stacking is the natural way in which graphene layers are stacked inside graphite. In this case, for $N=2M+1$ layers the low-energy electronic dispersion shows $2M$ parabolic energy bands and 1 linear band\cite{partoens_graphene_2006}. On the other hand, for the ABC case the energy band structure for small $k$ disperses as $|k|^N$  such that in the limit of large number of layers ($N$) the lower-energy band becomes flat over a large region in k-space  \cite{guinea_electronic_2006}.  This suppression of the kinetic energy results in low-energy surface states localized in the outer layer with a diverging density of states around the \textit{K} point. 

In order to describe superconducting correlations in multilayer graphene we use a tight-binding mean-field Hamiltonian, which in Nambu space can be written as follows:	
\begin{eqnarray}
\ham=\sum_{ \substack{<l,m> \\ <i,j>} } \vech{c_{l\up}^{i\dagger}}{c_{l\down}^{i}}
\mat{\hat{\ham}_{lm}^{ij}}{\Delta^i_l\delta_{ij}\delta_{lm}}{\Delta_l^{i\ast}\delta_{ij}\delta_{lm}}{-\hat{\ham}_{lm}^{ij\dagger}
}
 \vecv{c_{m\up}^{j}}{c_{m\down}^{j\dagger}} 
\label{ham}
\end{eqnarray}
where the summation, $\langle i,j \rangle$, is done over nearest neighbors within each layer while the summation, $\langle l,m \rangle$, is done for  adjacent layers. The superconducting order parameter $\Delta_l^i$ corresponds to s-wave singlet-pairing which is non-zero only in the contacts above the top-most graphene layer. Normal states are described by the Hamiltonian $\ham^{ij}_{lm}$, which within the simplest single-orbital tight-binding model is expressed as: 
\begin{eqnarray}
\label{ham0}
\hat{\ham}^{ij}_{lm}=[-t_0(1-\delta_{i,j})-\mu_l\delta_{i,j}]\delta_{l,m}-t(\delta_{l,m+1}+\delta_{l,m-1}) \;\;\;
\end{eqnarray}
where $\mu_l$ is the chemical potential and nearest-neighbor sub-lattices A and B within the $l$-th layer and labeled as A$_l$
and B$_l$  are coupled by the hopping parameter 
$t_0$, while $t=0.1t_0$ represents the electron transfer between the interlayer neighboring atomic sites  A$_l$ and B$_{l\pm1}$ from the adjacent $(l\pm1)$-th layer. 
Both stacking configurations are defined according to the vertical symmetry along the z-axis as shown in
Fig.~\ref{layout}. While the vertical atomic arrangement of ABA shows that we have only one sub-lattice per layer (A$_l$ or B$_l$) participating in the interlayer coupling, both sub-lattices from each layer are directly coupled in the ABC stacking.  Despite the fact that both atomic configurations are different we assumed  for simplicity the same intra-layer and inter-layer hopping integrals to be $t_0$ and $t$, respectively for both cases. We define as dimer (non-dimer) atoms, the atoms that are coupled (not coupled) to adjacent layers.\\
\begin{figure}[ttt]
\includegraphics[width=\columnwidth]{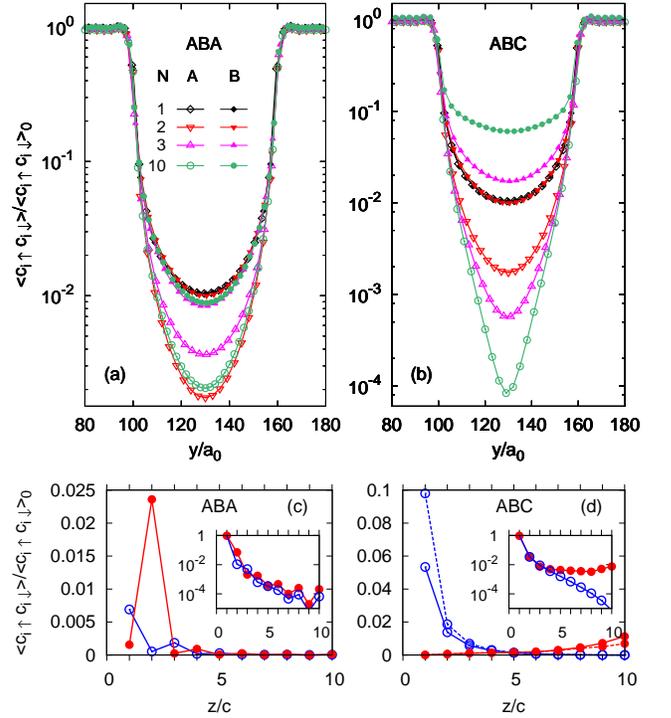}	
\caption{ \label{ycorr} (Color online) Absolute value of the pair correlation along the line $(x,y,z)=(W/2,y,1)$ in a single-layer graphene JJ and in the top-most layer in (a) ABA and (b) ABC multilayer graphene JJs with $N$=2, 3 and 10 layers. $a_0$=$3a/2$ where $a$ is the intra-layer C-C distance.  Pair correlation function at the middle point of the junction along the $\hat{z}$ vertical axis, $(x,y,z)=(W/2,L/2,z)$, for (c) ABA and (d) ABC multilayer graphene JJs with $N$=10 layers. The integer index, $z/c$, labels the different layers where $z/c=1$ correspond to the top-most layer and open(close) circles correspond to A(B) sub-lattice (see Fig.~\ref{layout}). Dashed lines in (d) correspond to the results obtained for small doping $\mu=0.0014t$. The insets in (c) and (d) show the vertical profile of the pair correlation at the middle point right underneath the superconducting contacts. 
}
\end{figure}
Since these types of structures involve a large number of atoms (each graphene layer is considered to have hundreds of thousands of atomic sites) numerical calculations are performed  by implementing the
Chebyshev-Bogoliubov-de-Gennes method \cite{covaci_efficient_2010, covaci_superconducting_2011} instead of performing an exact diagonalization of the Hamiltonian shown in Eq.~(\ref{ham}). In this way, we can numerically obtain an approximation of the Gorkov- Green's functions by a superposition  of Chebyshev polynomials as follows:
\begin{eqnarray}
\label{greenseries}
\bar{G}_{ijlm}^{1\alpha}(\tilde{\omega})=\frac{-2i}{\sqrt{1-\tilde{\omega}^2}}\left[\sum_{n=0}^N
a_{ijlm}^{1\alpha}(n)e^{-in\arccos(\tilde{\omega})}\right],
\end{eqnarray}
where the expansion coefficients for the normal ($\alpha=1$) and anomalous ($\alpha=2$), components of
the 2$\times$2 Green function are defined respectively as
\cite{covaci_efficient_2010}:
\begin{eqnarray}
\label{normal}
a^{11}_{ijlm}(n) = \la c_{l\up}^i\left|T_n(\ham)\right|c_{m\up}^{j\dagger}\ra \\
\label{anomalous}
a^{12}_{ijlm}(n) = \la c_{l\down}^{i\dagger}\left|T_n(\ham)\right|c_{m\up}^{j\dagger}\ra^{\ast}
\end{eqnarray}
where $T_n(x)=\cos[n \arccos(x)]$ is the Chebyshev polynomial of order $n$, which  satisfies
the following recurrence relation: $T_{n+1}(x)=2xT_n(x)-T_{n-1}(x)$.
Physical quantities, like the density of states, the pair correlation function and the Josephson current, can be easily determined once the Green function is known: $N_l^i(\omega)=-\frac{2}{\pi}\text{Im}{G}_{iill}^{11}(\omega)$, $\la c_{l\up}^i
c_{l\down}^i \ra = \frac{i}{2\pi}\int^{E_c}_{-E_c}{G}_{iill}^{12}(\omega)[1-2f(\omega)]d\omega$ and $J_{ij}^l = -\frac{1}{\pi}\int \text{Im}[ it_{ij}{G}_{ijll}^{11}(\omega)-it_{ij}^*{G}_{ijll}^{11*}(\omega)]f(\omega)d\omega$ respectively.
Once the Hamiltonian has been normalized according to $\ham\rightarrow \tilde{\ham}=(\ham-\eins b)/a$, where the rescaling
factors are $a=(E_{max}-E_{min})/(2-\eta)$ and $b=(E_{max}+E_{min})/2$, with $\eta>0$ being a small number, the expansion coefficients
can be obtained
through an iterative procedure involving a successive  application of the Hamiltonian on iterative vectors.
For more details about this numerical procedure we refer the reader to  Ref.~[\onlinecite{covaci_efficient_2010}].
Since the most consuming computational effort comes from the  sparse Hamiltonian matrix multiplications on iterative vectors, the performance is dramatically increased by implementing a parallel algorithm on graphical processing units (GPUs).
We are therefore able to solve efficiently for the electronic properties of multilayer graphene structures containing typically several millions of atoms by achieving significant speedup through computations on GPUs.

We next present the results of our simulations. In Fig.~\ref{ycorr} we show the pair correlation  profile along a line $(x,y,z)=(W/2,y,1)$ for single and multilayer graphene JJs with $N=$ 2, 3 and 10. In the case of  multilayer graphene the profile corresponds to the top-most graphene layer.  As expected, in single layer graphene JJ, the pair amplitude decays exponentially in the non-superconducting link in the same way for the A and B sub-lattice. However, a sub-lattice polarization is observed in the behavior of the pair amplitude for the multilayer graphene junctions where the Cooper pair diffusion across the junction has different decay characteristics in the A and B sub-lattice.

Previous self-consistent calculations performed by us revealed a similar sub-lattice polarization in the pair correlation function along a bilayer graphene JJ \cite{munoz_tight-binding_2012}. It can be observed in Fig.~\ref{ycorr} that such a polarization of the pair amplitude in the surface depends strongly on the stacking configuration for the multilayer cases with $N\geq3$. 
While no relevant differences in the pair depletion at the surface are observed for ABA stacking between the different values of \textit{N}, a peculiar dependence on $N$ is observed for the ABC case where the leaking distance is highly sensitive to the flatness of the lower energy band, i.e. to the number of layers $N$. 
Note in Fig.~\ref{ycorr}(b) that the interlayer coupled B sub-lattice in the top-most layer shows a suppression in the pair correlation while this is enhanced in the A sub-lattice which does not have a direct neighbor in the adjacent layer. In this way, and different from ABA, pair leaking in ABC is found to be larger than in the case of a single layer graphene JJ.
\begin{figure}[tt]
\includegraphics[width=\columnwidth]{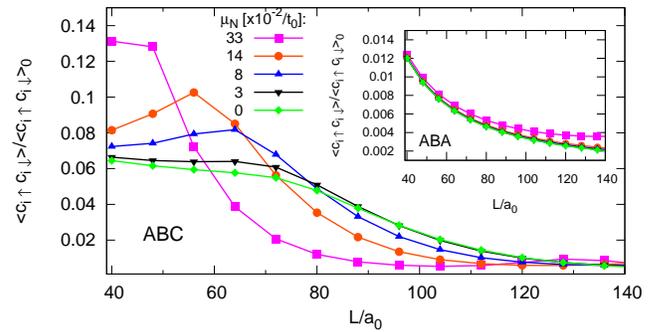}	
\caption{\label{dopef} (Color online) Amplitude of the pair correlation at the middle point $(x,y,z)=(W/2,L/2,1)$  as a function of the junction length $L$ for different values of the doping $\mu_N$ for the B sub-lattice in the top-most layer of the ABC and ABA (inset) multilayer graphene JJ containing $N=10$ layers.
 }
\end{figure}

Complementary results are shown in Figs.~\ref{ycorr}(c) and (d) where we plot the pair amplitude along the vertical axis $\hat{z}$ at the midpoint of the junction, $(x,y,z)=(W/2,L/2,z)$,  as a function of the integer index $z/c$, which labels the different layers starting from the top-most layer with $z/c=1$ (see Fig.~\ref{layout}), for both ABA and ABC stacking. Notice that the diffusion perpendicular to the contacts happens quite differently for  ABC and ABA stacking. For ABC stacking the pair leaking through the layers decays exponentially for the B sub-lattice while at the same time increases slowly for the A sub-lattice. 
For ABA stacking the pair amplitude for both dimer and non-dimer sites decays with a coherence length much smaller than the one observed in ABC (for the A sub-lattice). A particular behavior is found for ABA in the non-dimer site at the $(z/c=2)$-layer, where a rise of the pair amplitude is observed. This fact can be explained from a density of states point of view because it is well know that the local density of states in non-dimer sites is enhanced in the bulk while being suppressed at the surface. We can therefore conclude that the vertical leaking distance is larger in ABC than in ABA. 

Since previous results are performed at the Dirac point, i.e. for $\mu=0$, we have included in Fig.~\ref{ycorr}(d) an additional case for which the junction is slightly doped by setting a chemical potential $\mu=1.4\times 10^{-3}t_0$. 
These results are shown as dashed lines in Fig.~\ref{ycorr}(d) where we see that the pair correlations are enhanced in the A sub-lattice of the upper layers while being suppressed in the B sub-lattice in the lower layers. Such a high sensitivity of the pair correlation  on the chemical potential results from the sharp peak found in the LDOS for the surface sites of ABC multilayer graphene. The LDOS calculations (not shown here) demonstrate that energy position of this peak is slightly shifted by the coupling of the multilayer graphene with the highly-doped superconducting leads. Therefore, a slight doping will reposition the peak at the Fermi level and contribute to a large modification of the pair correlation.

The pair amplitude for the region underneath the superconducting contacts is shown in the insets of Figs.~\ref{ycorr} (c) and (d) as a function of the integer index $z/c$. In this case, an exponential decay is observed for both stacking orders, with the difference that for dimer-sites in ABC stacking this decay is much faster.     
\begin{figure}[ttt]
\includegraphics[width=\columnwidth]{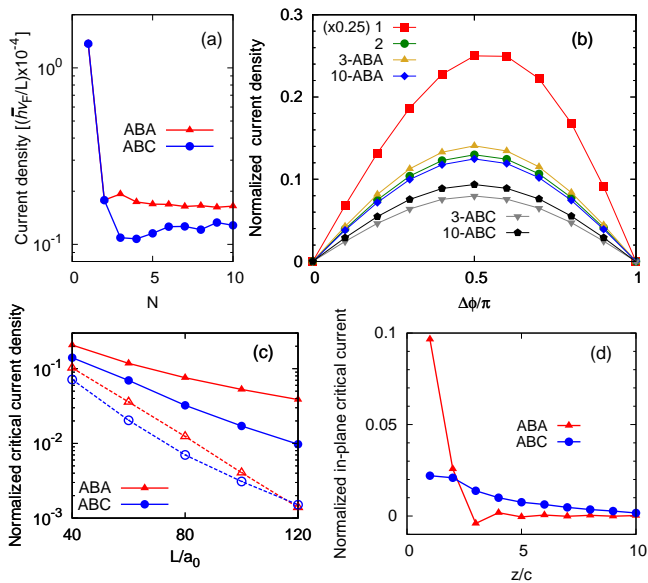}	
\caption{\label{curr} (Color online) (a) Josephson current as a function of the total number of layers, $N$ (b) Current density-phase relation for single ($N$=1), bilayer ($N$=2), trilayer ($N$=3) and multilayer ($N$=10) graphene JJs. Both stacking are considered. (c) Critical current density as a function of the junction lenght $L$ for $T$=0K (solid lines) and $T$=10K (dashed lines) and (d) $z$ rezolved in-plane critical current through the different graphene layers for ABC and ABA multilayer graphene JJs with $N$=10. Note that the current densities in panels (b)-(d) are normalized by the maximum Josephson current obtained for  single layer graphene, as shown in panel (a).
 }
\end{figure}
The dependence of the pair correlation at the midpoint of the junction corresponding to the A-sub-lattice in the top-most layer is shown in Fig.~\ref{dopef} as a function of the length of the junction for different values of the chemical potential $\mu$ . 
The $L$-dependence for ABC is notably different from the usual exponential decay observed in the ABA case (see inset Fig.~\ref{dopef}). Each doping level exhibits a maximum for different values of $L$. The unusual behavior is a consequence of the electron-hole asymmetry induced in the junction by the contact. Since very low values of doping in the normal region shifts the sharp LDOS peak localized at the surface, this has a strong effect on the decay of the pair amplitude. We found that a maximal effect is achieved when the value of the doping coincides with the energy position of the peak from the undoped case for each junction length. As this effect is coming from the influence of the leads on the surface state, there is a small dependence of this optimal doping on $L$.    

Since there are remarkable differences in the pair diffusion between ABC and ABA multilayer graphene it is expected that the superconducting current behaves differently in the two cases. In order to induce a zero-voltage super-current we set a phase difference, $\Delta\phi$, between the left and right superconducting contacts. We observe that the same phase difference is kept along the vertical axis $\hat{z}$ between two regions underneath the contacts which are away from the junction. However, near and inside the multilayer junction, the gradient of the phase along both the $\hat{z}$ and $\hat{y}$ directions varies in a different way in the two stacking configurations. This shows that the supercurrent is finite both within the intra-layers and between the inter-layers.

In Fig.~\ref{curr}(a) we plot the current density integrated over $\hat{z}$ as a function of the number of layers for the two stacking configurations. Notice that the largest current density was obtained for single layer graphene while for the cases with $N\geq2$ the current is suppressed. Interestingly, the current is always higher for ABA than ABC. In Fig.~\ref{curr}(b) we plot the current-phase relation and observe the same dependence on the stacking configuration. For $N\geq2$ the current-phase relation has the conventional $sin(\Delta\phi)$ dependence for short junctions. The length dependence of the current density is shown in Fig.~\ref{curr}(b) for two temperatures, 0 and 10K. Clearly, the current decays exponentially with the junction length. This is in contrast to the experimental finding from Ref.[\onlinecite{kanda_dependence_2010}], where a linear length dependence was uncovered. We believe that the main reason for this discrepancy comes from the different experimental setup, which considers a bottom gate that dopes  only the lower layers. 

In order to further elucidate these discrepancies we plot in Fig.~\ref{curr}(d) the $\hat{z}$ profile of the current for both stacking and find that the surface current is highly dominant in ABA stacking for which most of the current is observed in the two upper layers. This also explains the very weak dependence of the ABA current on the number of layers (see Fig.~\ref{curr}(a)). 
Alternatively, for ABC stacking the current is much more spread throughout the whole multilayer also explaining the stronger dependence of the current on the number of layers. In addition, for ABC stacking the larger the number of layers, the flatter the low energy band will be, which in turn will have an effect on the supercurrent by enhancing it.

In conclusion, by using a numerical tight-binding  approach for multilayer graphene contacted by two superconducting electrodes, we showed how the Cooper pairs diffuse both perpendicular and across the junction. We found that the proximity effect as well as the induced supercurrent are strongly dependent on the stacking configuration of the multilayer structure. For both ABA and ABC stacking we observe a polarization of the pair amplitude between dimer and non-dimer sites. This effect is enhanced in ABC stacking due to the peculiar flat band at the Fermi level which is localized at the surface.
Interlayer pair leaking is found to decay exponentially with a vertical-leaking distance larger in ABC than in ABA stacking. Despite the fact that the proximity effect is enhanced in ABC we found that the induced-current is larger in ABA but most of the current flows through the first two surface layers as opposed to ABC where the current is spread throughout the whole structure. We are therefore proposing that future experimental setups should use all the gates on the same side of the multilayer in order to take advantage of the surface currents. 

This work was supported by the Flemish Science Foundation (FWO-Vl) and the Methusalem funding of the Flemish Government.

\def\urlprefix{}
\def\url#1{}


%

\end{document}